\documentstyle[epsfig,longtable]{aipproc}

\begin{document}
\title{Stochastic Gravitational Radiation from Phase Transitions}

\author{Arthur Kosowsky$^*$, Andrew Mack$^*$ and Tinatin Kahniashvili$^{\dagger}$}
\address{$^*$Department of Physics and Astronomy, Rutgers University\\
Piscataway, New Jersey 08854-8019\\
$^{\dagger}$Abastumani Astrophysical Observatory, Tbilisi, 
Republic of Georgia}

\maketitle

\begin{abstract}
A stochastic background of gravitational radiation from cosmological
processes in the very early Universe is potentially detectable. 
We review the gravitational radiation which may arise from
cosmological phase transitions, covering both bubble collisions
and turbulence as sources. Prospects for detecting a direct
signal from the electroweak phase transition or other cosmological
sources with a space-based
laser interferometer are discussed. 
\end{abstract}

\section*{Introduction}

Scientific and design goals for gravitational radiation detectors
have been driven primarily by point sources of radiation like
binary compact objects. While the distribution and properties of such
sources will undoubtedly be of cosmological interest, the most
interesting gravitational wave signals for cosmology are likely to be
stochastic backgrounds produced in the early Universe. 
Since the Universe is opaque to electromagnetic radiation at redshifts
greater than $z=1300$ due to Thomson scattering off free electrons,
gravitational radiation offers the only hope of ``imaging'' the
earliest epochs of the Universe when it was dominated by high-energy
fundamental physics processes.

Several possible stochastic sources have been considered. The
best motivated are cosmic defects like strings or textures; quantum
fluctuations in the fields driving inflation; and phase transitions.
More speculative ideas include string-theory inspired scenarios for
cosmic evolution at times prior to the apparent initial singularity
(so-called ``pre-big-bang'' cosmology) or explosive preheating
at the end of inflation.
Here we review the stochastic background of
gravitational radiation expected from first-order phase transitions.
New results for gravitational radiation 
from turbulence, which could be produced in a first-order phase
transition, are included. The final section considers the
detectability of various stochastic sources, with an emphasis on
the electroweak phase transition.
A thorough review of stochastic gravitational radiation, with
extensive material about detector phenomenology, overviews
of a variety of sources, and a comprehensive bibliography,
has recently appeared \cite{mag00}.

\section*{General Considerations}
The time evolution of cosmological gravitational radiation follows
directly from the linearized Einstein equations,
\begin{equation}
{\ddot h}_{ij}({\bf k},\eta) 
+ 2{{\dot a}\over a}{\dot h}_{ij}({\bf k},\eta) 
+ k^2 h_{ij}({\bf k},\eta)
= 8\pi Ga^2 \Pi_{ij}({\bf k},\eta),
\label{hij_evol}
\end{equation}
where ${\bf k}$ is a comoving wave vector, $\eta$ is conformal time,
overdots are derivatives with respect to $\eta$, 
$a$ is the scale factor of the Universe,
$h_{ij}$ is the tensor metric perturbation, and $\Pi_{ij}$ is the
tensor piece of the stress-energy tensor. We will also make use of the
Hubble parameter, $H=(1/a) da/dt$ and its present value $h_{100}$ in units
of 100 km/s/Mpc, and throughout we use natural units
with $\hbar = c = k_B = 1$. The right side of
Eq.~(\ref{hij_evol}) is the source term for generating gravitational radiation,
while the left side describes the free propagation of the radiation,
including damping due to the expansion of the Universe. Since each
comoving ${\bf k}$-mode is independent, the physical wavelength of a
given gravitational wave increases linearly with the scale factor $a$,
just as with electromagnetic radiation. 

Solutions to the homogeneous equation with $\Pi=0$ are easily
obtained. For a radiation-dominated universe, $a\propto\eta$ and
\begin{mathletters}
\begin{equation}
h_{ij}\propto j_0(k\eta),\qquad y_0(k\eta)
\label{h_rad}
\end{equation}
while for the matter-dominated era, $a\propto\eta^2$ and
\begin{equation}
h_{ij}\propto {j_1(k\eta)\over k\eta },\qquad {y_1(k\eta)\over k\eta},
\label{h_mat}
\end{equation}
\end{mathletters}
where $j_l$ and $y_l$ are the usual spherical Bessel functions. It is
straightforward to show that in both cases, the amplitude of the
gravitional waves scales as $a^{-1}$. This result also follows from
the stress-energy tensor for gravitational radiation, whose energy
density is given by \cite{mtw73}
\begin{equation}
\rho_{GW} = {1\over 32\pi G}\left\langle 
{\partial h_{ij} \over \partial t} 
{\partial h^{ij} \over \partial t}
\right\rangle
\label{rho_gw}
\end{equation}
where the average is over many wavelengths. So the energy density is
proportional to $h_c^2 f^2$, where $h_c$ is the characteristic
amplitude of the metric perturbation $h_{ij}$ and $f$ is the
(physical) frequency of the wave. The energy density scales like
$a^{-4}$ with the expansion of the Universe, while the frequency
scales like $a^{-1}$. It follows that $h_c\propto a^{-1}$ as derived
above from the evolution equation. Note this is different from the
amplitude scaling of the electric and magnetic
fields in an electromagnetic wave, which drop
off like $a^{-2}$. 

In practice, only very large stress-energy sources contribute
significantly to the stochastic gravitational wave amplitude. If a
particular source acts for a period of time shorter than the
Hubble time $H^{-1}$, then the expansion of the Universe can be neglected
during the time the source is active and the resulting gravitational
wave amplitude can be conveniently calculated in flat space. Then the
entire subsequent evolution of the gravitational waves is simply a
reduction of amplitude and frequency by the factor 
\begin{equation}
{a_*\over a_0} = 8.0\times 10^{-14} 
\left( 100\over g_*\right)^{1/3}
\left(1\,\,{\rm GeV}\over T_*\right), 
\label{scaleratio}
\end{equation}
where
$a_*$ and $T_*$ are the scale factor and temperature
at the time the source is active, $g_*$ is the total number of
relativistic degrees of freedom at the time of the source, and
$a_0$ is the scale factor today. This expression is valid as long as
the evolution of the Universe since the source has been adiabatic.
If a source exists for a time
interval long compared to the Hubble time, then the expansion of the
Universe may impact the source evolution and the full evolution
equation, Eq.~(\ref{hij_evol}), must be used.

A stochastic background of gravitational waves can be
characterized by $\Omega_{\rm GW}(f)$, 
its energy density per frequency octave
in units of the critical density
$\rho_c = 3H^2/8\pi G$. From this, we follow Thorne \cite{tho87}
in defining a
convenient characteristic amplitude at frequency $f$ as
\begin{equation}
h_c(f)\equiv 1.3\times 10^{-18}
\left[ \Omega_{\rm GW}(f) h_{100}^2\right]^{1/2} 
\left(1\,\,{\rm Hz}\over f\right).
\label{h_c}
\end{equation}
We will calculate characteristic energy densities in gravitational
wave backgrounds and convert to characteristic amplitudes with this
relation. 

\section*{Model Phase Transitions}

The most interesting potential source of a cosmological gravitational
wave background is an early-Universe phase transition.  Phase
transitions very likely occurred in the early Universe. The standard
model of particle physics provides two: the electroweak phase
transition at $T\simeq 100$ GeV, at which the electroweak symmetry was
spontaneously broken, and the QCD phase transition at $T\simeq 150$
MeV, at which chiral symmetry was broken. If the standard
model is actually the result of the breaking of a larger gauge
symmetry (i.e. a Grand Unified Theory), then a GUT phase transition at
an energy scale of $T\simeq 10^{16}$ GeV is also likely.  Other more
speculative phase transitions involving the breaking of hypothesized
additional symmetries have also been discussed.

The details of a particular phase transition, in particular its order,
latent heat, and dynamics, are determined by the effective potential
driving the phase transition. In a given particle theory, calculation
of an effective potential is in general a difficult, non-perturbative
problem of finite-temperature field theory. For purposes of
gravitational wave signals, we consider a generic, model first-order
phase transition described by several physical parameters; a specific
phase transition corresponds to some set of model parameters.

As the Universe cools due to its expansion, phase transitions can
occur if a new state with lower energy density becomes physically
possible. Assume this occurs at some characteristic temperature scale
$T_*$. The energy density difference between the two phases is roughly
the vacuum energy density, $\rho_{\rm vac}$. If the two phases are
separated by a significant potential energy barrier (greater than the
characteristic thermal energy density at temperature $T_*$), then the
transition must proceed via the nucleation of new-phase bubbles via
quantum or thermal processes. We define $\alpha\equiv\rho_{\rm
vac}/a_RT_*^4$, the ratio of vacuum energy density to thermal energy
density ($a_R$ is the radiation constant).  
Once a bubble of new phase is nucleated, the potential
energy difference between the two phases exerts an outward force on
the walls of the bubble, causing it to expand. Hydrodynamic forces
will tend to resist the bubble expansion, which will quickly attain
some equilibrium velocity $v$. We neglect possible instabilities which
might distort the bubble's spherical shape and redistribute energy.
As the bubbles expand, some fraction $\kappa$ of the available vacuum
energy is converted to the kinetic energy of the expanding bubble
walls, while the rest goes into heat.

Once multiple bubbles have been nucleated in some region of the
Universe, they will expand until eventually their walls meet and they
merge together, converting the entire region to the lower-energy
phase. The energy stored up in the bubble walls is eventually
dissipated into heat via hydrodynamical turbulence.  If $\alpha$ is
large enough, the kinetic energy of the bubble walls will represent a
significant fraction of the entire energy density of the Universe, and
will have $v$ a significant fraction of $c$. Note that for $T_* >
10^5$ K, the energy density of the Universe is dominated by radiation,
so the sound speed will be $c_s = c/\sqrt{3}$, the relativistic
limit. As soon as the spherical symmetry of an individual bubble is
broken due to collisions with other bubbles, gravitational radiation
will be produced; the high energy densities and velocities involved
make bubble collisions a potentially potent source
\cite{wit84,hog86,tw90}.

The characteristic frequency of the gravitational radiation is
determined by the duration of the phase transition, corresponding to
the percolation time of the biggest bubbles with the largest
energies. A simple bubble nucleation model takes an exponential bubble
nucleation rate per unit volume $\Gamma = \Gamma_0\exp(\beta t)$
\cite{tww92}. This form is motivated by the general observation that
$\Gamma$ will typically be the exponential of some nucleation action,
and the time dependence of the nucleation action can be approximated
as linear at the time of the phase transition. For such a nucleation
rate, the duration of the phase transition is roughly
$\tau\simeq\beta^{-1}$. The characteristic length scale is just
$\beta^{-1} v$, the size of the largest bubbles at the end of the
phase transition. In general, $\beta\simeq 4\ln(m_{\rm Pl}/T_*) H_*
\simeq 100 H_*$, where $H_*$ is the Hubble parameter at the time of
the phase transition \cite{tww92}. Thus the characteristic wavelength
of gravitational radiation at the time of the phase transition 
is perhaps a percent of the Hubble radius at that time.

A simple model of a first-order phase transition is comprised of the
quantities defined above: the characteristic time scale $\beta^{-1}$;
the vacuum energy $\rho_{\rm vac}$; the bubble expansion velocity $v$;
the temperature of the phase transition $T_*$ or equivalently the
ratio of vacuum energy to thermal energy $\alpha$; and the efficiency
factor $\kappa$.

The gravitational radiation resulting from the violent collision of
bubbles in a first-order phase transition has been calculated in
detail \cite{ktw92,kt93,kkt94}. It was demonstrated that the
radiation source is primarily the bubble walls, which for can be
treated in the thin-wall approximation if the bubble walls propagate
as detonation fronts. The traceless portion of the stress-energy
tensor is nonzero only at the position of the spherically expanding
bubble walls, and the amplitude of the stress-energy tensor follows
simply from the amount of vacuum energy liberated by a particular
bubble's expansion.  Then the radiation from many randomly nucleated
bubbles in a sample volume can be computed numerically. The result is
a stochastic background of radiation with energy density
\begin{equation}
\Omega_{\rm GW}(f) h^2 \simeq 1.1 \times 10^{-6} \kappa^2
\left({H_*\over\beta}\right)^2 \left({\alpha\over 1+\alpha}\right)^2
\left({v^3\over 0.24 + v^3}\right)\left({100\over g_*}\right)^{1/3},
\label{collide_omega}
\end{equation}
peaking at a characteristic frequency 
\begin{equation}
f_{\rm max} \simeq 5.2 \times 10^{-8}\,{\rm Hz}
\left(\beta\over H_*\right) \left(T_*\over 1\,{\rm GeV}\right)
\left(g_*\over 100\right)^{1/6},
\label{collide_fmax}
\end{equation}
with characteristic amplitude
\begin{equation}
h_c(f_{\rm max})\simeq 1.8\times 10^{-14}\kappa
\left({\alpha\over\alpha+1}\right)
\left(H_*\over\beta\right)^2
\left(1\,{\rm GeV}\over T_*\right)
\left({v^3\over 0.24+v^3}\right)^{1/2}
\left(100\over g_*\right)^{1/3}.
\label{collide_hc}
\end{equation}
At frequencies lower than the characteristic peak frequency, the
energy density per frequency octave increases like $f^{2.5}$ 
while above this frequency it drops off more slowly \cite{kt93}. 
For very strong
phase transitions, the energy spectrum of gravitational radiation has
a long tail at high frequencies, due to the presence of numerous small
bubbles with significant energy densities.  These results hold for a
strongly first-order phase transition where the bubbles expand
supersonically as detonation fronts. For weaker transitions with
deflagration (subsonic) bubble walls, the dynamics are more complex,
but the resulting radiation is expected to be much smaller.

Note that the frequency range of the proposed LISA space-based
interferometer, $10^{-5}$ to $10^{-1}$ Hz, is particularly well-suited
to probing the electroweak phase transition at an energy scale of 100
GeV: $\beta/H_*$ is generically around 100, putting the frequency for
the peak of the gravity wave power spectrum somewhere between
$10^{-3}$ and $10^{-4}$ Hz. The strength of the electroweak phase
transition is an open question; this issue is discussed below in the
concluding section.

\section*{Model Turbulence}

Substantial gravitational radiation can also be produced by a period
of turbulence in the early Universe \cite{kkt94}. The following
analysis \cite{mkk01} is completely general, 
without reference to a particular
source of turbulence, although the most likely source is a first-order
phase transition: once the bubbles of low-temperature phase begin to
collide with each other, they stir up the primordial plasma on a
characteristic length scale corresponding to the size of the largest
bubbles in the transition. The gravitational radiation produced by the
turbulence will be in addition to that from the bubbles given above.

We construct a simple model of isotropic cosmological turbulence as
follows: at some particular temperature $T_*$ and corresponding
enthalpy $w_* = \rho_* + P_*$ the turbulence commences, with energy
input into the Universe on a largest length scale $L_S$. In general,
the energy input will be a complicated process; we make the simple
idealization that the turbulence will last for a time $\tau$ and that
the turbulent energy will be distributed as a stationary Kolmogoroff
spectrum with
\begin{equation}
E(k) \equiv {1\over w}{d\rho_{\rm turb}\over dk} \simeq
{\bar\varepsilon}^{2/3} k^{-5/3},
\label{kolmogoroff}
\end{equation}
where $\bar\varepsilon$ is the energy density dissipation rate per
unit enthalpy \cite{my75}.
The turbulent energy injected at the scale
$L_S$ will be written as $\epsilon\rho_{\rm vac}$, with $\epsilon$ an
efficiency factor.  This energy cascades to smaller scales until it is
dissipated by viscous damping at a scale determined by the kinematic
viscosity $\nu$ of the fluid. The damping scale $L_D$ can be obtained
from the Reynolds number, which can be approximated as
\begin{equation}
{\rm Re} = \left(L_S\over L_D\right)^{4/3} 
\simeq {2\over 3}\left(L_S\over 2\pi\right)^{4/3}
\left({\epsilon\rho_{\rm vac}\over \nu^3\tau w}\right)^{1/3},
\label{reynolds}
\end{equation}
if the turbulent source lasts for a long time compared to the
dynamical time scale of the turbulence on the scale $L_S$. 
The critical Reynolds number for the development of a
Kolmogoroff spectrum of turbulence is
around 2000; generally this number is easily exceeded for
early-Universe phase transitions. 
If the source lasts for a time short compared to the dynamical
time at the scale $L_S$ (as is likely with the electroweak phase
transition \cite{kkt94}), then fully developed turbulence never appears,
but we argue on physical grounds that, as far as
gravitational wave production is concerned, this case can be treated
like fully-developed turbulence which lasts for one
dynamical time on the scale $L_S$ \cite{mkk01}.

In analogy with the first-order
phase transition calculation, the model turbulence is completely
described by the physical quantities defined above: the characteristic
time scale $\tau$, which we write as $\beta H_*^{-1}$; the energy
density which goes into turbulent motions, $\epsilon\rho_{\rm vac}$;
the characteristic length scale $L_S$, which we write as $\gamma
H_*^{-1}$; the temperature of the phase transition $T_*$, which
determines the enthalpy density $w_*$; and the kinematic viscosity
$\nu$, which fixes the scale $L_D$ at which the turbulence is
dissipated into heat.

Given this model, the turbulent stress-energy tensor can be
constructed. The starting point is the Fourier-space expression
\begin{equation}
T_{ij}({\bf k}) = {w\over (2\pi)^3}\int d{\bf q}\, u_i({\bf q}) u_j({\bf k-q})
\label{Tij_turb}
\end{equation}
where ${\bf u}$ is the relativistic velocity of the fluid at a 
given point and $w$ is the fluid enthalpy, assumed to be
independent of position. The source 
for gravitational radiation can
be obtained via a projection tensor \cite{dfk00},
\begin{equation}
\Pi_{ij}({\bf k}) = \left(P_{il} P_{jm} - {1\over 2} P_{ij}P_{lm}\right)
T_{lm}({\bf k})
\label{Pi_ij}
\end{equation}
where $P_{ij}\equiv \delta_{ij} - {\bf\hat k}_i{\bf\hat k}_j$ 
is a projector onto the transverse plane.
The connection to isotropic turbulence is obtained via the
two-point correlation function
\begin{equation}
\left\langle u_i({\bf k}) u_j^*({\bf k'})\right\rangle
= (2\pi)^3 P_{ij} P(k) \delta({\bf k}-{\bf k'}) 
\label{uiuj}
\end{equation}
which holds for any statistically isotropic and homogeneous 
velocity field. For Kolmogoroff turbulence, the power
spectrum is
\begin{equation}
P(k)\simeq \pi^2{\bar\varepsilon}^{2/3} k^{-11/3}.
\label{power_kol}
\end{equation}
After a lengthy calculation which combines these pieces,
the resulting gravitational radiation
spectrum can be estimated as \cite{mkk01}
\begin{equation}
h_c(f)\simeq 7\times 10^{-14} \gamma^{5/3}\beta^{1/3}
\left({\epsilon\rho_{\rm vac}\over w_*}\right)^{2/3}
\left(100\over g_*\right)^{1/3}\left({1\,{\rm GeV}\over T_*}\right)
\left({f_S\over f}\right)^{4/3},
\label{h_turb}
\end{equation}
with the lowest frequency $f_S$ corresponding to 
the comoving scale $L_S$, 
\begin{equation}
f_S = 1.6 \times 10^{-7}\,{\rm Hz}\, \gamma^{-1}
\left({T_*\over 1\,{\rm GeV}}\right)
\left(g_*\over 100\right)^{1/6},
\label{f_turb}
\end{equation}
using Eq.~(\ref{scaleratio}).  The spectrum of the gravitational
radiation extends to the highest frequency corresponding to the
dissipation length scale. This can be approximated, but it suffices to
note that the Reynolds number is large in the early Universe, so the
range of frequencies will span at least a factor of several
hundred. Turbulence can be just as efficient at producing
gravitational radiation as the first-order phase transition that
spawned it. In cases where the phase transition is only weakly
first-order, the turbulence may be the dominant source.

Turbulence has the possibility of producing gravitational radiation
via another mechanism. If any small seed magnetic field is present at
the onset of turbulence, a turbulent dynamo will amplify the field
exponentially until it saturates at equipartition with the turbulent
kinetic energy. In Kolmogoroff turbulence, the dynamical time on a
given scale is shorter at smaller scales, so the diffusion scale
magnetic fields will saturate first. Typically, the smallest scale
dynamical time is at least 100 times shorter than the largest scale
dynamical time, so the magnetic seed field on the smallest scales will
be amplified by at least 100 e-foldings.  We can crudely approximate
the resulting magnetic field as isotropic with a power spectrum that
has the Kolmogoroff form from the smallest scale up to some saturation
scale, and then an exponential drop up to the largest scale.

The magnetic fields generated by turbulence are potentially
interesting because they also generate gravitational radiation.
Roughly, the equipartition magnetic fields are as efficient as the
turbulence at making gravitational waves, with the crucial difference
that the magnetic fields last much longer. The turbulence damps out on
the turbulence time scale, while the magnetic fields remain until
matter-radiation equality. The resulting gravitational radiation
energy density can be approximated as that from the turbulence times
an additional factor of $\ln(z_*/z_{\rm eq})$, where $z_*$ is the
redshift of the turbulence and $z_{\rm eq}$ is the redshift of
matter-radiation equality. For the electroweak phase transition, this
factor is around 25, giving a large increase in the gravitational wave
amplitude. This mechanism would yield a distinctive power spectrum of
gravitational waves: it peaks at a frequency corresponding to the
largest scale on which the dynamo saturates, drops off like a power
law at smaller scales, and drops off exponentially at larger scales.
The major question about this mechanism is whether an adequate seed
field exists. The plasma in the very early Universe is strongly in the
MHD regime (collision time short compared to dynamical time) and has a
very high conductivity, making it difficult for any seed field
generation mechanism to work. It has been suggested that the expanding
bubbles in the electroweak phase transition will produce adequate seed
fields \cite{soj97}, while thermal seed fields have also been
advocated \cite{taj92}. We are currently investigating gravitational
radiation from turbulent-dynamo magnetic fields in detail.

\section*{Detectability of Stochastic Sources}

Phase transitions are almost sure to have occurred in the early
Universe; in particular, electroweak symmetry breaking at $T\simeq
100$ GeV and the QCD phase transition at $T\simeq 1$ GeV follow from
our current understanding of the standard model of particle physics.
However, only phase transitions which are first order are likely to
have caused large enough stress-energy fluctuations to leave behind a
significant gravitational wave background. In the standard model, both
the electroweak and QCD phase transitions appear to be
second-order. On the other hand, no high-energy theorist believes that
the standard model is the final story. Once additional symmetries and
fields are allowed, as with supersymmetry,
the strength of the electroweak phase transition
can be greatly enhanced. Belief in a strongly first-order electroweak
phase transition is motivated by electroweak baryogenesis \cite{tro99},
which requires significant departure from thermodynamic equilibrium
and thus a strongly first-order phase transition (but see
\cite{tg99}).  Baryogenesis at the electroweak energy scale is a
natural possibility, since at higher energies, sphaleron processes
which violate baryon number occur rapidly enough to erase any net
baryon number. The very existence of matter in the Universe
may be connected to a detectable gravitational wave background;
we are currently studying this issue. 

Inflationary gravitational radiation may be detectable with LISA,
although it is difficult to come up with sensible models which give
signals much above the anticipated LISA sensitivity \cite{lid94}.  The
amplitude of the gravitational radiation at the present horizon scale
is limited by the COBE measurement of large angular scale fluctuations
in the microwave background temperature, and general considerations of
slow-roll inflation argue that the power spectrum of gravitational
radiation must drop as the scale decreases. Detection of such a
background would be extremely important, because it would reflect the
energy scale at which inflation occurred.

Cosmic defect sources, while well-motivated, offer less hopeful direct
detection prospects.  At this point, measurements of large-scale
structure and the microwave background likely rule out any cosmic
defects as the primary mechanism for structure formation in the
Universe \cite{abr99}. On the other hand, defects are a generic
product of symmetry breaking. The gauge group of the standard model of
particle physics is SU(3)${}_C\times$SU(2)${}_L\times$U(1)${}_Y$,
which has a U(1) subgroup. A topological theorem states that the
breaking of any larger group to a subgroup containing a U(1) factor
will result in one-dimensional defects, i.e. cosmic strings. Thus the
production of cosmic strings is a generic feature of any cosmology
incorporating breaking of a grand-unified (GUT) symmetry to the
standard model.  (although if the GUT transition is inflationary, the
number density of cosmic strings will be exponentially suppressed,
making them completely irrelevant). Even if strings or other defects
do not drive structure formation, they could conceivably produce
detectable amounts of stochastic gravitational radiation \cite{vv85}.

Other cosmological sources of gravitational radiation, such as an
early string phase of cosmic evolution \cite{lwc00} or complex field dynamics
during reheating after inflation \cite{preheat}, 
offer detection possibilities, but
these sources are based on more speculative underlying physics.  Until
further theoretical advances put them on more secure footing, they are
perhaps best considered examples of the flavor of theories which might
give an unanticipated signal.

In summary, potential cosmological sources of a stochastic background
of gravitational radiation are more speculative than the compact
binary point sources towards which LIGO is largely oriented. But
several sources are well motivated, and the potential scientific
payoff is spectacular. In particular, a direct probe of electroweak
symmetry breaking with the LISA space-based interferometer is a
distinct possibility, as is the direct detection of gravitational
waves from inflation. All conjectured gravitational wave sources in
the early Universe are intimately connected with fundamental physics
at energy scales between 100 GeV and $10^{16}$ GeV. For all but the
lowest portion of this range, gravitational radiation represents the
{\it only} direct probe of physics available with currently envisioned
technology. It also seems likely that stochastic backgrounds offer a
greater chance at serendipidous discovery than point sources, since
the evolution of the very early Universe is determined by physics
which is partly unknown, while point sources are probably known
astrophysical objects. We strongly encourage designers of
gravitational radiation detectors to make stochastic background
sensitivity a primary design consideration.

\medskip

This work has been supported by NASA's Astrophysics Theory Program.
T.K. has been partially supported by the National Research Council's
COBASE program. A.K. is a Cotrell Scholar of the Research Corporation.

\end{document}